\documentclass[a4paper,conference]{IEEEtran}
\ifCLASSINFOpdf
\else
\fi
\usepackage{epsfig}
\usepackage{graphicx}
\usepackage{amsmath}
\usepackage{amssymb}
\usepackage{multirow}
\usepackage{adjustbox}
\usepackage{csquotes}
\usepackage{subcaption}
\usepackage{comment,balance}
\usepackage{array, caption, tabularx,  ragged2e,  booktabs}
\usepackage{cite}

\begin{document}
%
\title{Extremely Low-light Image Enhancement with Scene Text Restoration}


%
\author{\IEEEauthorblockN{Pohao Hsu\IEEEauthorrefmark{1}\textsuperscript{\textsection},
Che-Tsung Lin\IEEEauthorrefmark{2}\textsuperscript{\textsection},
Chun Chet Ng\IEEEauthorrefmark{3}\textsuperscript{\textsection},
Jie-Long Kew\IEEEauthorrefmark{3}, 
Mei Yih Tan\IEEEauthorrefmark{3}, \\
Shang-Hong Lai\IEEEauthorrefmark{1}\IEEEauthorrefmark{5},
Chee Seng Chan\IEEEauthorrefmark{3} and
Christopher Zach\IEEEauthorrefmark{2}}
\IEEEauthorrefmark{1}National Tsing Hua University, Hsinchu, Taiwan\\
\IEEEauthorrefmark{2}Chalmers University of Technology, Gothenburg, Sweden\\
\IEEEauthorrefmark{3}Universiti Malaya, Kuala Lumpur, Malaysia \\
\IEEEauthorrefmark{5}Microsoft AI R \& D Center, Taipei, Taiwan}


\maketitle
\begingroup\renewcommand\thefootnote{\textsection}
\footnotetext{These authors contributed equally to this work.}
\endgroup

\begin{abstract}
Deep learning-based methods have made impressive progress in enhancing extremely low-light images - the image quality of the reconstructed images has generally improved. However, we found out that most of these methods could not sufficiently recover the image details, for instance, the texts in the scene. In this paper, a novel image enhancement framework is proposed to precisely restore the scene texts, as well as the over- all quality of the image simultaneously under extremely low-light images conditions. Mainly, we employed a self-regularised attention map, an edge map, and a novel text detection loss. In addition, leveraging synthetic low-light images is beneficial for image enhancement on the genuine ones in terms of text detection. The quantitative and qualitative experimental results have shown that the proposed model outperforms state-of-the-art methods in image restoration, text detection, and text spotting on See In the Dark and ICDAR15 datasets.
\end{abstract}

%
\IEEEpeerreviewmaketitle

\section{Introduction}

\label{sec:intro}

\begin{figure}[!t]
	\centering
	\begin{subfigure}{.21\linewidth}
  \centering
 \includegraphics[keepaspectratio=true, scale = 0.11]{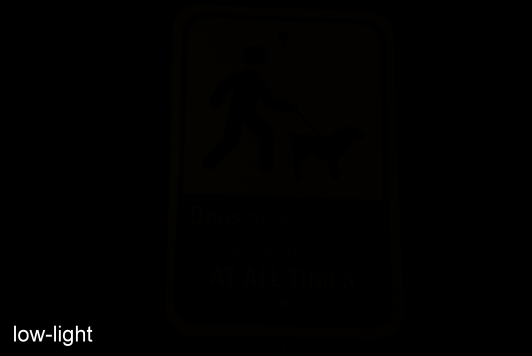}
 \includegraphics[keepaspectratio=true, scale = 0.11]{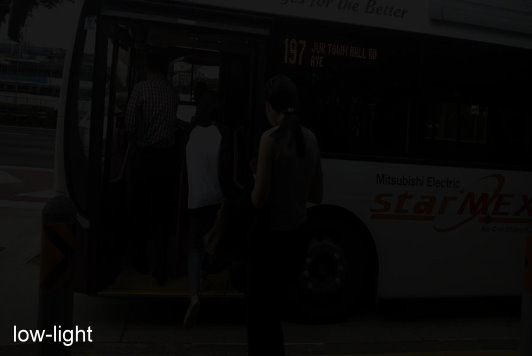}
 \caption{}
\end{subfigure}
\hfill
\begin{subfigure}{.21\linewidth}
  \centering
\includegraphics[keepaspectratio=true, scale = 0.11]{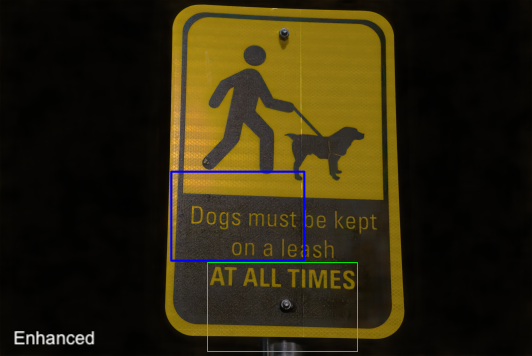}
	\includegraphics[keepaspectratio=true, scale = 0.11]{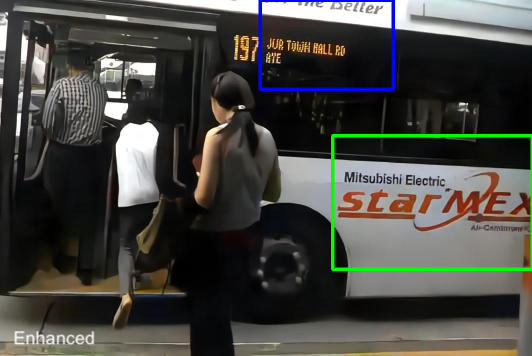}
	 \caption{}
\end{subfigure}
\hfill
\begin{subfigure}{.21\linewidth}
  \centering
\includegraphics[keepaspectratio=true, scale = 0.11]{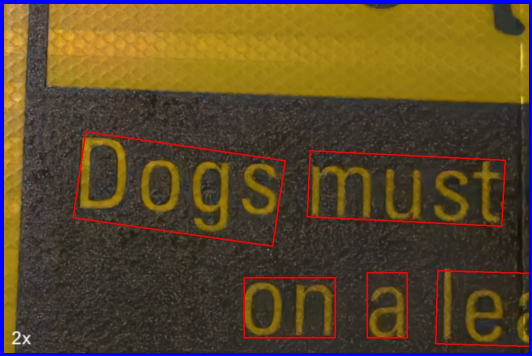}
	\includegraphics[keepaspectratio=true, scale = 0.11]{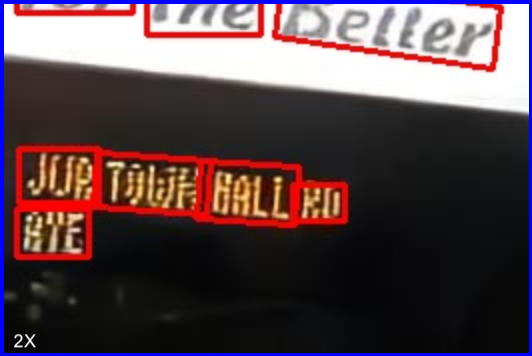}
	 \caption{}
\end{subfigure}
\hfill
\begin{subfigure}{.21\linewidth}
  \centering
  \includegraphics[keepaspectratio=true, scale = 0.11]{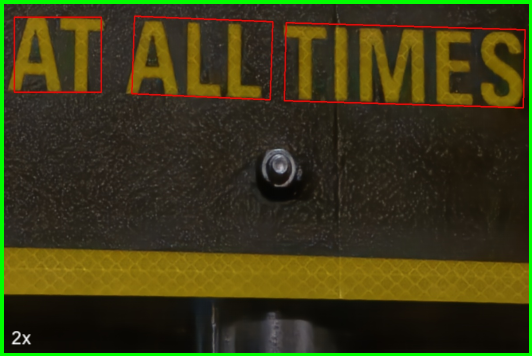}
	\includegraphics[keepaspectratio=true, scale = 0.11]{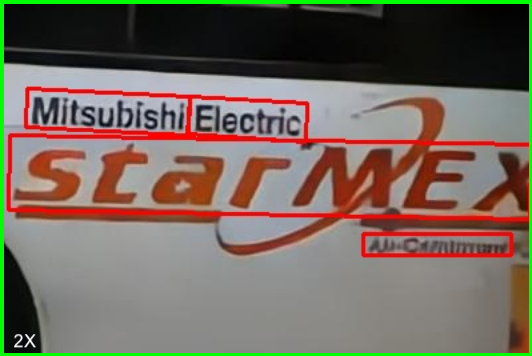}
	 \caption{}
\end{subfigure}
	\caption{From left to right: (a) Original low-light images; (b) Enhanced results with our proposed method; (c-d) Zoomed-in (2x) of the blue and green bounding box. It is obvious that the texts are clearly visible with sharp edges.}
	\label{fig:overall_results}
\end{figure}

Image enhancement under extremely low-light conditions is very challenging as under such an extreme scenario, a large amount of information is usually not available, as shown in Figure \ref{fig:overall_results}(a). At the same time, these images are also highly susceptible to noise due to demosaicing in the image sensing pipeline. Although it is possible to capture a better image with either a larger aperture or a longer exposure time, the images might then suffer from being overexposed or blurred due to improper camera settings or/and object motions.



Recently, many low-light enhancement algorithms have been proposed. Traditional approaches~\cite{Pizer1987AdaptiveHE,elik2011ContextualAV,lee2013contrast,jobson1997properties,Jobson1997AMR} usually aim at restoring the statistics of the original images to that of the normal or natural ones; while deep learning-based methods~\cite{Tao2017LowlightIE,Tao2017LLCNNAC,Lore2017LLNetAD,Gharbi2017DeepBL,Wei2018DeepRD,CycleGAN2017,Jiang2019EnlightenGANDL,isola2017image,zhu2020eemefn,Shi2019LowlightIE} mainly aim to learn the mapping between low-light images and brighter ones via regression. 

Overall, the quality of these low-light images has been improved to a certain extent. However, we found out that the degree of visibility or readability of contextual information, for instance, the scene texts in these enhanced low-light images has never been explicitly discussed or stressed. As an example, empirically, we notice that current low-light image enhancement models generally suffer from losing finer details, especially in the text regions. That is to say, although the existing methods can achieve moderate image quality scores, e.g., peak signal-to-noise ratio (PSNR) and structural similarity (SSIM) in overall, they seem to fail in the downstream tasks, such as text detection and text spotting in our case as shown in Figure \ref{fig:sony_fig}-\ref{fig:icdar_fig} (b)-(h). On the other hand, with the success of deep neural networks, text detection~\cite{Baek2019CharacterRA,wang2019efficient} and text recognition~\cite{trba2019,Aster2019} models have made significant progress. They can even work on extremely low-light images to small extent (see Figure \ref{fig:icdar_fig} (a)), but the results can be significantly improved (see Figure \ref{fig:icdar_fig} (i)) if the images are better restored in terms of both the overall image quality and the local delicate text features.

In this paper, we are inspired to address this problem by focusing on restoring texts in scene images captured under extremely low-light conditions and the overall image quality simultaneously. Diving into more details, below is the
summary of contributions of this paper:

\begin{itemize}
\item Firstly, we proposed an image enhancement framework capable of improving both the low-light image quality (in overall) and the scene text simultaneously through a novel text detection loss. It enables the enhanced images to preserve finer low-level details such as character edges and shapes without compromising the overall quality of the image, leading to a successful text detection (see Figure \ref{fig:overall_results}, Tables \ref{table:PSNR_all}-\ref{table:det_ocr_all}).
\item Secondly, in terms of datasets, we (i) annotated the texts in the real low-light dataset - See In the Dark (SID) Sony~\cite{Chen2018LearningTS}, and (ii) created a synthetic low-light dataset based on the commonly used ICDAR15~\cite{Karatzas2015ICDAR2C} dataset. Both datasets will be made publicly available upon acceptance. 
\item Thirdly, extensive experiments have shown that our proposed model can achieve the best scores in terms of text detection and spotting tasks on the enhanced low-light images in both See In the Dark and ICDAR15 datasets. We have also shown that our image enhancement model trained with additional synthetic low-light images can achieve better results on genuine ones (Table \ref{table:mix_data_ablation}). 
\end{itemize}

\section{Related Works}
Generally, low-light image enhancement can be categorized into two main approaches: traditional methods and deep learning-based ones. Over the years, traditional Histogram Equalization (HE) and its variations~\cite{Pizer1987AdaptiveHE,elik2011ContextualAV,lee2013contrast} have been widely studied. To overcome their limitations, Retinex-based algorithms, such as SSR~\cite{jobson1997properties}, MSR~\cite{Jobson1997AMR}, SIRE~\cite{Fu2016AWV}, LIME~\cite{Guo2017LIMELI}, and BIMEF~\cite{Ying2017ABM} were proposed and assume that an image can be decomposed into illumination and reflectance. Recently, deep learning-based image denoising and enhancement tasks have achieved significant improvements. DCNN~\cite{Tao2017LowlightIE} proposes a joint framework with a CNN-based denoising module. LLCNN~\cite{Tao2017LLCNNAC} designs a special module to utilize multi-scale feature maps. LLNet~\cite{Lore2017LLNetAD} uses a stacked-sparse denoising autoencoder to identify signal features and adaptively enhances and denoises the image. Inspired by bilateral grid processing and local affine color transforms, HDRNet~\cite{Gharbi2017DeepBL} predicts image transformation by learning to make local, global, and content-dependent decisions. Retinex-Net~\cite{Wei2018DeepRD} combines deep learning and retinex theory, and adjusts illumination for enhancement after image decomposition. 

Furthermore, the recent successes of GANs~\cite{Goodfellow2014GenerativeAN} have also attracted attention from the low-light image enhancement community because GANs have proven successful in image synthesis and translation. Specifically, Pix2pix~\cite{isola2017image} can provide visually plausible images in the target domain given paired training data. However, it is challenging to obtain paired images in many practical applications. As such, cycle-consistency introduced by CycleGAN~\cite{CycleGAN2017} had opened up the possibility of performing unpaired image-to-image translation. To overcome the complexity of CycleGAN, EnlightenGAN~\cite{Jiang2019EnlightenGANDL} proposed an unsupervised one-path GAN structure which includes a global discriminator and a local one, a self-regularized perceptual loss fusion, and an attention mechanism. EEMEFN~\cite{zhu2020eemefn} proposed an edge enhancement module to enhance the initial image generated by the multi-exposure fusion (MEF) module; however, the model focuses on raw images only. Retinex-GAN~\cite{Shi2019LowlightIE} presented a method combining the Retinex theory and GAN.

One of the nearest works to us is possibly Xue et al. \cite{xue2020arbitrarily} that proposed a low light text detector based on spatial and frequency feature fusion to enhance the fine details of low light image. However, in contrast to their work, we propose an image enhancement framework to jointly improve the image and text quality as a whole, as illustrated in Figure \ref{fig:architecture}.

\section{Proposed Method}
\begin{figure*}[!t]
	\centering
	\includegraphics[width=\linewidth]{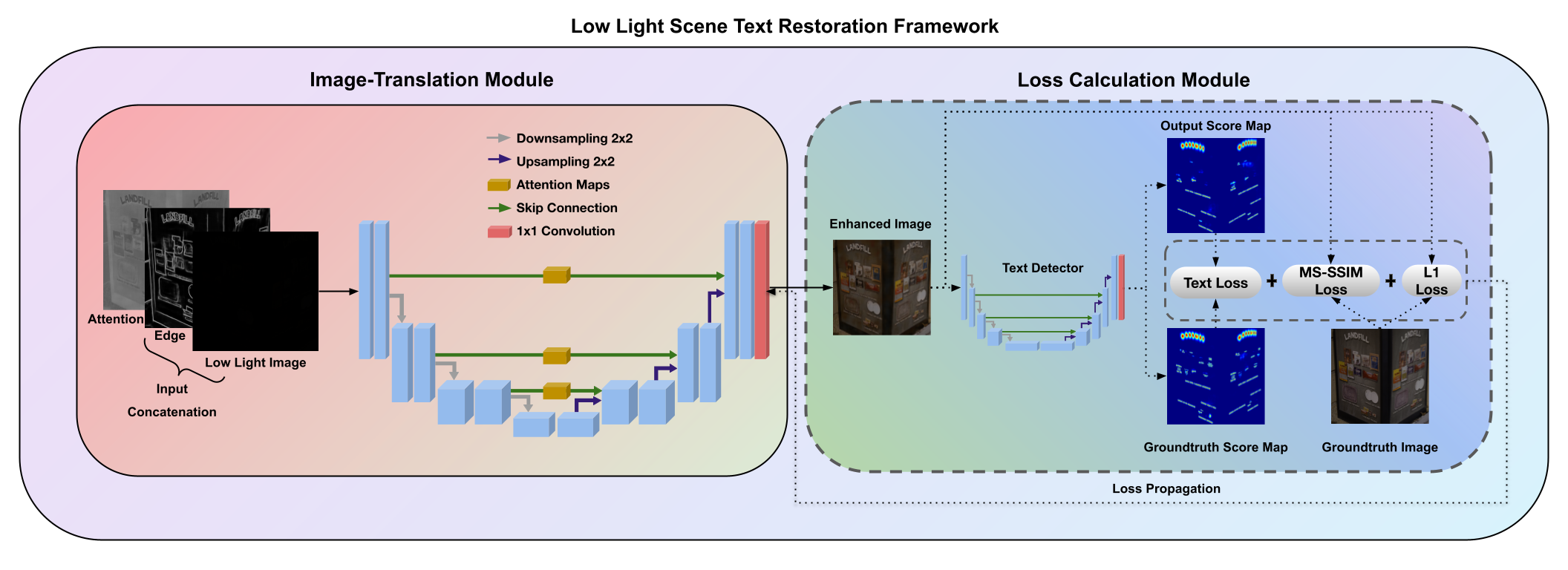}
	\vspace{-3mm}
	\caption{Illustration of the overall architecture of our proposed low-light scene text restoration framework. Starting with the input concatenation of low-light image and edge map, the core enhancement network is guided by (1) an attention module through the multiplication of feature map with attention map, and followed by (2) the calculation of novel text detection loss where the model focuses on scene text regions. The dashed lines show how the text detection loss is formulated and propagated back to the main enhancement network so that the image enhancement and text restoration tasks can be trained simultaneously.}
	\label{fig:architecture}
\end{figure*}


This paper introduces a novel framework that can simultaneously enhance the scene text and overall image quality under extreme low-light conditions. Our proposed framework consists of two modules, as illustrated in Figure \ref{fig:architecture}. The first module is the image-translation module that consists of a U-net which takes as input a combination of a low-light image, an edge map and an attention map, while the second module is the loss calculation module which is only involved in the training phase and consists of an ${\ell_1}$ loss, MS-SSIM loss, and our novel text detection loss.

\subsection{Image-Translation Module}
Herein, we adopt the U-net generator with refinements. i.e., edge map and self-regularized attention map which will be described next into the U-net. As such, our U-net generator concatenates the low-light image \textit{$I$} and the edge map \textit{$E$} as a 4-channel input with self-regularized attention map \textit{$S$}: 
\begin{equation}
	I' = \mathcal{F}(I, E, S).
\end{equation}

\subsubsection{{\bf Edge map}}
RCF edge detector~\cite{Liu2019RicherCF} can be directly employed to predict edges from the low-light images. The idea is that ${\ell_1}$ or mean squared error loss tends to blur sharp edges and other fine image details. Therefore, leveraging edge information can recover abundant textures and sharp edges and is helpful in the following downstream tasks, such as text detection. Technically, the RCF edge detector is pretrained on the BSDS500 dataset~\cite{arbelaez2010contour} based on the VGG16 architecture. The network consists of five stages that will extract multiple levels of edge features containing each layer's essential fine details. Then, these features from all different stages are combined by a fusion layer. In this paper, instead of directly using RCF to predict edges, which might lead to blurred edges and artifacts in the dark regions, we trained another U-Net to produce the edge maps.


\subsubsection{{\bf Self-regularized attention map}}
Low-light image enhancement can easily cause the output image to be over or under-exposure. Recently, EnlightenGAN~\cite{Jiang2019EnlightenGANDL} proposed an easy-to-use attention mechanism called self-regularized attention to handle this issue by focusing on the dark areas in the images. Technically, the self-regularized attention map \textit{$S$} is attained by taking the illumination channel \textit{$Y$} of the input RGB image, normalizing it to [0,1], followed by \textit{$1 - Y$} (element-wise difference). Herein, we also found this trick helpful in extremely low-light image enhancement. In our work, the attention map is consistently re-scaled through max pooling layers and element-wisely multiplied with the feature maps in the encoding part. Then, these re-weighted features are passed to the corresponding feature maps in the decoding part at different scales as skip-connections.

As a whole, our U-Net consists of 9 convolutional blocks and a 1x1 convolutional layer on top. Each block consists of two 3x3 convolutional layers followed by LeakyReLu. Each block is connected with a 2x2 max pooling layer at the downsampling step. We also apply max pooling on the attention map simultaneously. At the upsampling step, we extract the feature map with 2x2 transposed convolution and concatenate it with the corresponding feature map from the downsampling step, which passes through the attention maps. 


\subsection{Loss Calculation Module}
As aforementioned, previous image enhancement methods suffered from not being able to restore scene text regions sufficiently in the low-light images. Therefore, in this module, we introduced a text detection loss $\mathcal{L}_{text}$, as well as two other image enhancement loss ($\mathcal{L}_{SSIM_{MS}}$ and $\mathcal{L}_{\ell_1}$) as a joint loss function to simultaneously improve the overall image quality, as well as the scene text quality.

\subsubsection{{\bf Text detection loss}}
Intuitively, a well-restored scene text implies that we could obtain a very similar text detection results on the enhanced image and the ground truth. In this work, we employ CRAFT~\cite{Baek2019CharacterRA} as our text detector to effectively localize the text characters and then estimate the affinity between them for text detection. Technically, our CRAFT network architecture is based on VGG16 with batch normalization as the backbone. There are skip connections in the decoding part, which is similar to U-Net in aggregating low-level features. CRAFT model will predict two Gaussian heatmaps: (i) \textit{region score} and (ii) \textit{affinity score}. In the former, the \textit{region score} is the probability of the characters, and in the latter, the \textit{affinity score} represents the probability that adjacent characters are in the same word. As such, we propose the text detection loss \textit{$\mathcal{L}_{text}$} as: 
\begin{equation}
	\mathcal{L}_{text} = \| R(I') - R(I^{GT})\|_1,
\end{equation}
where $R(I')$ and $R(I^{GT})$ denote the region score maps of the enhanced image and the ground-truth image, respectively; while $w$ and $h$ denote the width and height of the region score map.

\subsubsection{{\bf Image enhancement loss}}
The multi-scale SSIM method was proposed in~\cite{Wang2003MultiscaleSS} for reference-based image quality assessment, focusing on the image structure consistency. An $M$-scale SSIM between $I'$ and $I^{GT}$ is given by
\begin{multline}
	SSIM_{MS}(I', I^{GT}) = [l_M(I', I^{GT})]^{\alpha} \cdot \\ \prod\nolimits^M_{j=1} [c_j(I', I^{GT})]^{\beta} [s_j(I', I^{GT})]^{\gamma},
\end{multline}
where $l_M$ is the luminance at \textit{M}-scale; $c_j$ and $s_j$ represent the contrast and the structure similarity measures at the $j$-th scale, respectively; while $\alpha$, $\beta$, and $\gamma$ are parameters to adjust the importance of the three components.

Inspired by ~\cite{Wang2003MultiscaleSS}, we adopted the multi-scale SSIM loss function in our work to enforce the image structure of the enhanced image $I'$ to be close to that of the ground-truth image $I^{GT}$:
\begin{equation}
	\mathcal{L}_{SSIM_{MS}} = 1 - {SSIM_{MS}}(I', I^{GT}).
\end{equation}

Additionally, in order to better enforce correctness at the low frequencies~\cite{isola2017image}, we also employ $\ell_1$ loss between $I'$ and $I^{GT}$ as:
\begin{equation}
	\mathcal{L}_{\ell_1} = \|I' - I^{GT}\|_1.
\end{equation}

As summary, the overall joint loss function to train our image enhancement model is given by:
\begin{equation}
	\mathcal{L}_{Total} = \omega_1 \mathcal{L}_{\ell_1} + \omega_2 \mathcal{L}_{SSIM_{MS}} + \omega_3 \mathcal{L}_{text},
	\label{eq2total}
\end{equation}
\noindent where $\omega$ is the hyperparameter.

\begin{table*}[!ht]
	\begin{center}
		\begin{tabularx}{\textwidth}{l c c c c c c c c}
			\hline
			& SRIE~\cite{Fu2016AWV} & LIME~\cite{Guo2017LIMELI} & BIMEF~\cite{Ying2017ABM} & RetinexNet~\cite{Wei2018DeepRD} & CycleGAN~\cite{CycleGAN2017} &  EnlightenGAN~\cite{Jiang2019EnlightenGANDL} & Pix2pix~\cite{isola2017image} & Ours\\
			\hline
			Sony & 12.59/0.104 & 13.87/0.135 & 12.87/0.110 & 15.49/0.368 & 15.34/0.453 &  14.59/0.426 & 21.07/0.662 & \bf{25.51}/\bf{0.716} \\
			Syn. ICDAR15 & 10.42/0.409 & 12.04/0.522 & 16.00/0.581 & 15.55/0.637 & 23.42/0.720 &  21.03/0.661 & 24.87/0.727 & \bf{28.41}/\bf{0.840} \\
			\hline
		\end{tabularx}
	\end{center}
	\caption{Quantitative evaluation of low-light image enhancement algorithms in terms of PSNR/SSIM.}
	\label{table:PSNR_all}
\end{table*}

\begin{table*}[!ht]
	\begin{center}
		\resizebox{\textwidth}{!}{
			\begin{tabular}{l|cc|cc|cc|cc}
				\hline
				\multicolumn{1}{c|}{\multirow{3}{*}{Model}} & \multicolumn{4}{c|}{Text Detection H-Mean} & \multicolumn{4}{c}{Two-Stage Text Spotting Case Insensitive Accuracy}\\ \cline{2-9} 
				\multicolumn{1}{c|}{} & \multicolumn{2}{c|}{Sony} & \multicolumn{2}{c|}{Syn. ICDAR 15} & \multicolumn{4}{c}{Syn. ICDAR 15}\\ \cline{2-9}
				\multicolumn{1}{c|}{} & CRAFT & PAN & CRAFT & PAN & CRAFT + TRBA & CRAFT + ASTER & PAN + TRBA & PAN + ASTER\\ \cline{1-9}
				Input & 0.057 & 0.026 & 0.355 & 0.192 & 0.114 & 0.118 & 0.062 & 0.067 \\ \hline
				SRIE~\cite{Fu2016AWV} & 0.133 & 0.076 & 0.465 & 0.423 & 0.127 & 0.144 & 0.117 & 0.122 \\
				LIME~\cite{Guo2017LIMELI} & 0.127 & 0.057 & 0.454 & 0.428 & 0.129 & 0.146 & 0.120 & 0.128 \\
				BIMEF~\cite{Ying2017ABM} & 0.136 & 0.079 & 0.450 & 0.411 & 0.124 & 0.138 & 0.123 & 0.122 \\
				RetinexNet~\cite{Wei2018DeepRD} & 0.115 & 0.040 & 0.374 & 0.325 & 0.090 & 0.096 & 0.069 & 0.076 \\
				CycleGAN~\cite{CycleGAN2017} & 0.090 & 0.053 & 0.428 & 0.458 & 0.119 & 0.144 & 0.122 & 0.138 \\
				EnlightenGAN~\cite{Jiang2019EnlightenGANDL} & 0.146 & 0.075 & 0.458 & 0.461 & 0.140 & 0.157 & 0.123 & 0.139\\
				Pix2pix~\cite{isola2017image} & 0.266 & 0.190 & 0.559 & 0.542 & 0.183 & 0.207 & 0.182 & 0.195\\
				Ours & \bf{0.324} & \bf{0.266} & \bf{0.623} & \bf{0.631} & \bf{0.193} & \bf{0.219} & \bf{0.197} & \bf{0.221}\\ \hline
				GT & 0.842 & 0.661 & 0.800 & 0.830 & 0.526 & 0.584 & 0.555 & 0.591\\ \hline
		\end{tabular}}
	\end{center}
	\caption{Quantitative evaluation of enhanced images of all image enhancement algorithms in terms of H-Mean for text detection, and we use case insensitive word accuracy as the main text spotting score. Scores in bold are the best of all.}
	\label{table:det_ocr_all}
    \vspace{-3mm}
\end{table*}

\section{Experimental Results}
\subsection{Experiment Setup}
{\noindent\bf Datasets.} Two public datasets, \textbf{SID}~\cite{Chen2018LearningTS} Sony and  \textbf{ICDAR15}~\cite{Karatzas2015ICDAR2C}, are employed in this work.
\textbf{SID} Sony was captured by Sony $\alpha$7S II contains 2697 short-exposure images and 231 long-exposure images at the resolution of 4240×2832. The exposure time of the short-exposure images was set to 1/30, 1/24, and 1/10 seconds. The corresponding long-exposure images were captured with 10 and 30 seconds. In our experiments, we convert the SID Sony images to 24-bit RGB format and manually label the scene text bounding boxes. As a result, there are 8210 and 611 labels for short-exposure and long-exposure images, respectively. We will publicly release the annotated dataset.
\textbf{ICDAR15} dataset was introduced in the ICDAR 2015 Robust Reading Competition for incidental scene text detection and recognition. It contains 1500 scene text images at the resolution of 1280x720. In this work, the brightness of each image in ICDAR15 is reduced to make it visually similar to SID Sony dataset, so that it forms paired images in the low-light image enhancement application.

{\noindent\bf Metrics.}
We compare our model with state-of-the-art methods in terms of PSNR and SSIM to measure the image quality. Also, we follow the common standard to use H-Mean as the evaluation metric for text detection. For text spotting, case insensitive word accuracy is used but lexicon is not employed so that we can study the raw impact of different sets of the enhanced images. Finally, the SID Sony dataset is only analyzed in terms of text detection because text recognition labels are not available. 

{\noindent\bf Implementation Details.}
We train our network for 4000 epochs using Adam optimizer~\cite{Kingma2015AdamAM}. The initial learning rate is set to 1e-4 which is decreased to 1e-5 after 2000 epochs. In each training iteration, we randomly crop a 512×512 patch, and apply random flipping together with rotation as data augmentation. The parameters in the loss function $(\omega_1, \omega_2, \omega_3)$ are set to $(0.85, 0.15, 0.425)$ empirically.

\subsection{Quantitative comparison}
We compare our model with traditional approaches, as well as deep learning-based. Table \ref{table:PSNR_all} reports PSNR and SSIM to indicate how much the image quality is improved. It is noticed that our method is able to obtain the highest scores on both SID Sony and ICDAR15 datasets, which signifies that our method can enhance the low-light images to be as close as possible to the original images. Table \ref{table:det_ocr_all} shows the scene text detection scores and text spotting accuracies of our model and other related works.

{\noindent\bf Scene Text Detection.}
We run two state-of-the-art scene text detectors (CRAFT~\cite{Baek2019CharacterRA} and PAN~\cite{wang2019efficient}) on the enhanced images. As shown in Table \ref{table:det_ocr_all}, our model achieves the highest H-Mean on SID Sony and synthetic ICDAR15 datasets.

{\noindent\bf Scene Text Spotting.}
In order to show that both detection and recognition tasks are highly correlated and better enhanced images will lead to a quantitatively higher text detection and text recognition accuracy at the same time, we carry out a two-stage text spotting experiment using the aforementioned detectors and two robust scene text recognizers (TRBA~\cite{trba2019} and ASTER~\cite{Aster2019}). In detail, we first select detection results that have Intersection over Union (IoU) greater than 0.5 when compared to the ground-truth bounding boxes. Then, we crop out the detected text regions before passing them to the text recognizers. As shown in Table \ref{table:det_ocr_all}, when both recognizers are fed with the text detection results of CRAFT~\cite{Baek2019CharacterRA} and PAN~\cite{wang2019efficient} respectively, it is evident that the recognition accuracy of the images enhanced by our model consistently outperforms the ones of other competing methods. This shows that our proposed method is able to enhance the overall images, as well as restoring the text regions where our results excel in terms two-stage text spotting. In contrary, current existing methods might introduce a lot of noises and artifacts due to the inferior image enhancement capability.



\subsection{Qualitative comparison}

\begin{figure*}[!ht]
	\includegraphics[width=\textwidth]{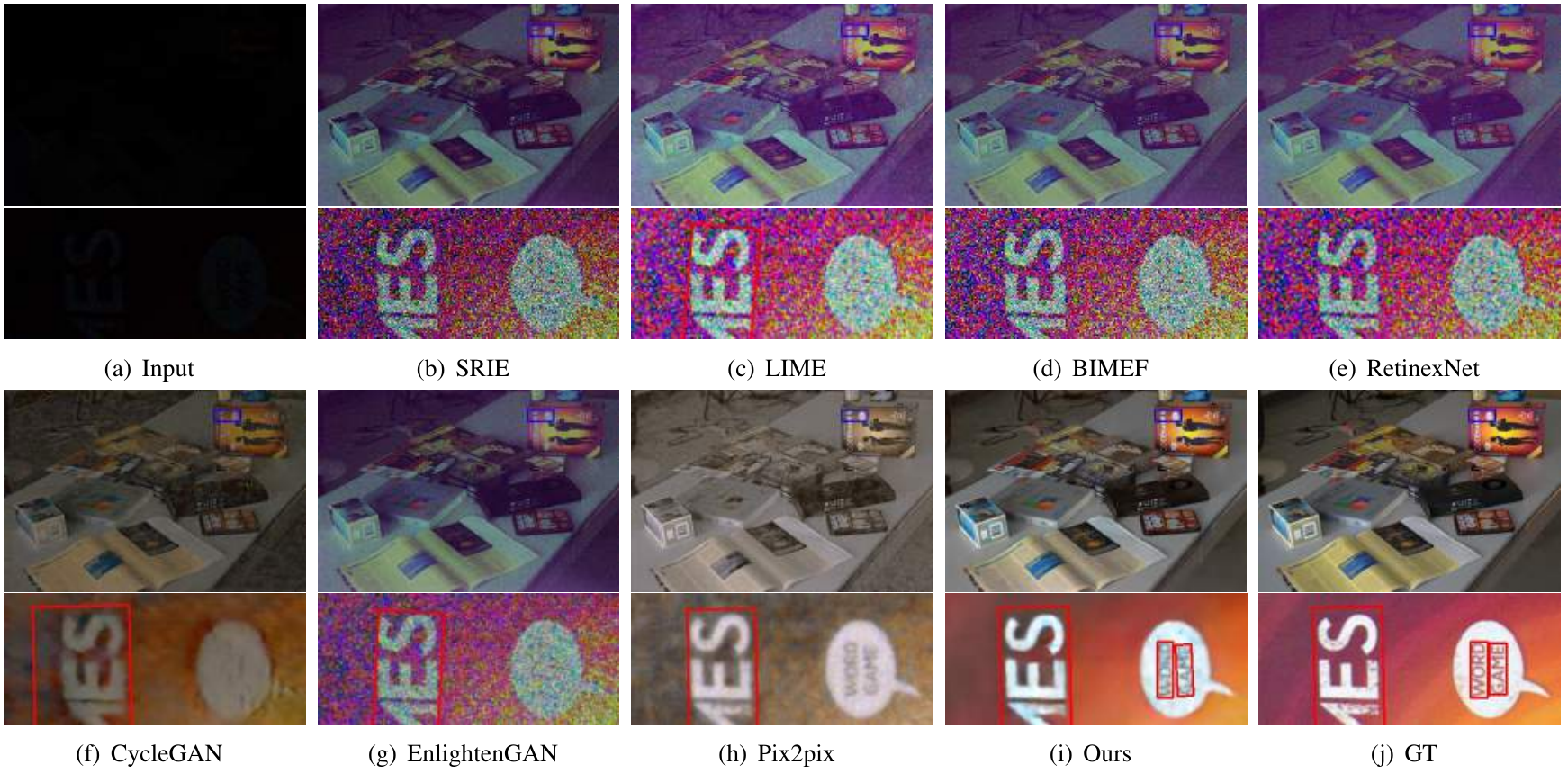}\\
	\vspace{-3mm}
	\caption{CRAFT's results (red boxes) of all methods on the SID Sony dataset, focused on the zoomed-in regions (blue boxes).}
	\label{fig:sony_fig}
\end{figure*}

\begin{figure*}
	\includegraphics[width=\textwidth]{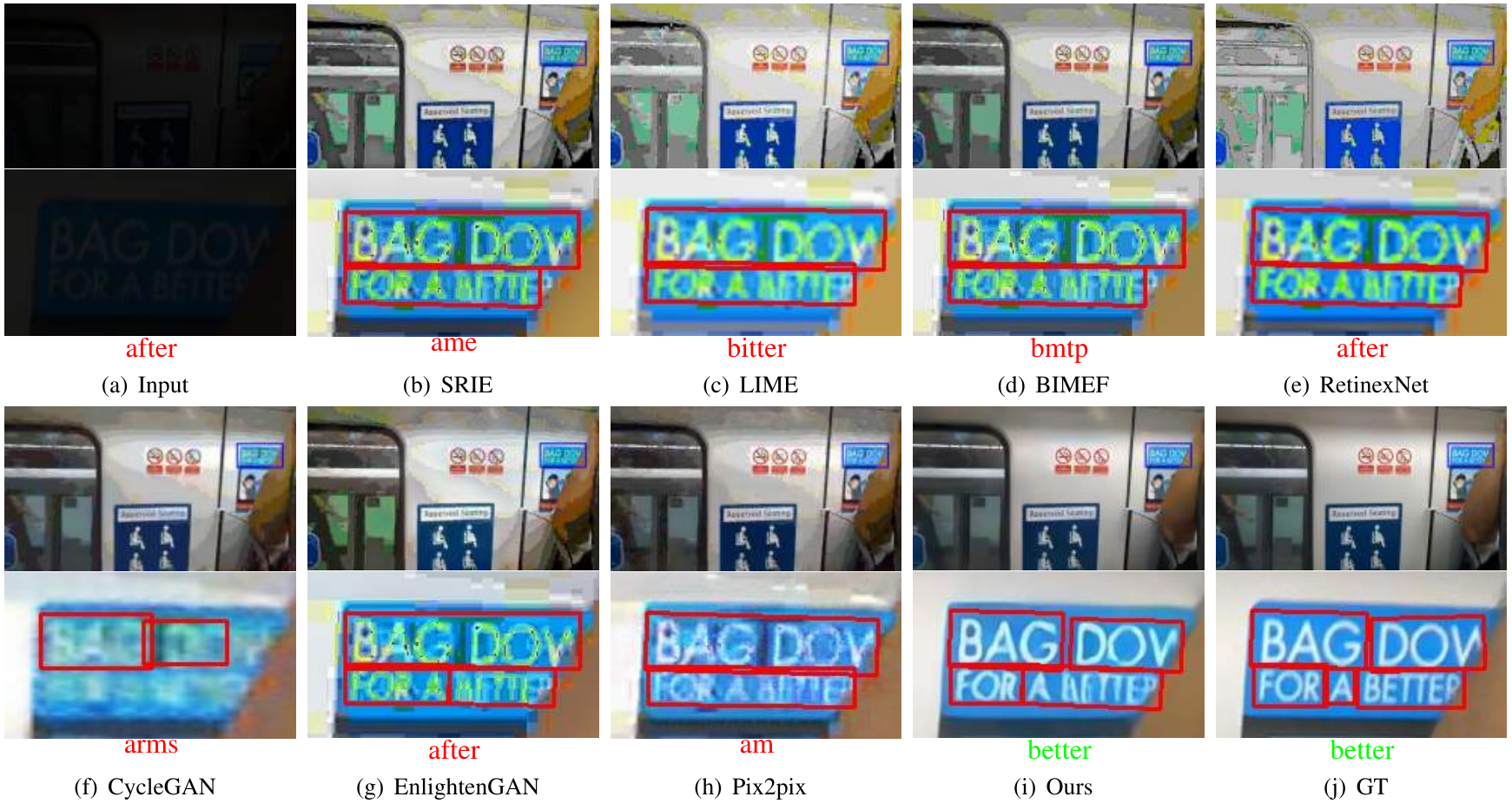}\\
	\vspace{-3mm}
	\caption{CRAFT's results (red boxes) of all methods on the ICDAR15 dataset, focused on the zoomed-in regions (blue boxes), followed by ASTER's results based on the ground-truth bounding box of \enquote{BETTER}.}
	\label{fig:icdar_fig}
\end{figure*}

Figure \ref{fig:sony_fig} and \ref{fig:icdar_fig} demonstrate the image enhancement results on SID Sony and synthetic ICDAR15 datasets, respectively. Figures \ref{fig:sony_fig}a and \ref{fig:icdar_fig}a are the original low-light input, and (b)-(i) are the images enhanced by SRIE~\cite{Fu2016AWV}, LIME~\cite{Guo2017LIMELI}, BIMEF~\cite{Ying2017ABM}, RetinexNet~\cite{Wei2018DeepRD}, CycleGAN~\cite{CycleGAN2017}, EnlightenGAN~\cite{Jiang2019EnlightenGANDL}, Pix2pix~\cite{isola2017image} and our proposed method. The last image in both figures is the ground truth image. We also show the zoomed-in details on selected scene text regions, overlaid with the text detection bounding boxes. 

It is noticed that the results obtained with LIME and RetinexNet contain overexposed noises, while SRIE and BIMEF can barely recover the images. Then, CycleGAN and Pix2pix have color distortion and could not restore details. Overall, these methods fail to recover the text regions to the degree that the text detector could detect them accurately.

{\noindent\bf Scene Text Detection.}
It can be noticed that the proposed model is able to enhance the low-light image and achieves the best performance in terms of scene text detection. For instance, in Figure \ref{fig:sony_fig}, only the images enhanced by our method can provide text detection results almost identical to the ground truth. Existing methods either miss certain words or fail to predict each word individually. This happens because only our method can produce enhanced images with less noise and finer details in terms of text detection.

{\noindent\bf Scene Text Spotting.}
Upon visualizing the recognition results of ASTER in Figure \ref{fig:icdar_fig} for the word \enquote{BETTER} in the ground truth, only our method is able to recognize the word correctly. While for other methods, recognition results of low-light input, RetinexNet, and EnlightenGAN are \enquote{after}. SRIE and LIME leads to the output of \enquote{ame} and \enquote{bitter} respectively. BIMEF yields \enquote{bmtp}, Pix2pix predicts \enquote{am}, and \enquote{arms} for CycleGAN. These results show that our method is able to enhance and restore the text at a more refined level, and it can preserve important low-level features such as edges and character strokes to achieve accurate text recognition results.
\subsection{Ablation Study}
To understand the effect of each component of our model, we conduct several ablation experiments by either adding or removing them one at a time. We employ U-Net architecture using $\ell_1$ loss as our baseline. In Table \ref{table:map_loss_ablation} and Table \ref{table:map_loss_ablation_without}, the first column shows the performance of the baseline, and the rightmost column is the result of our full model. We can observe that each component contributes to a higher text detection accuracy to a certain extent. For example, the self-attention module guides the network to focus on the dark areas instead of the bright ones, while the edge map demonstrates that the network can extract more information from the sharper edges and fewer artifacts. Multi-scale SSIM loss leads to a better text detection result because the enhanced images are closer to the original ones. The text detection loss encourages the model to restore the texts to the extent of being detectable by the text detector instead of solely improving the overall image quality. As a whole, each component contributes to the overall better text detection and recognition results for the enhanced extremely low-light images.


\begin{table}[!t]
	\begin{center}
		\begin{tabular}{c c c c c c}
			\hline
			baseline & w/ A & w/ A+E & w/ A+E+M  & w/ A+E+M+T (full model) \\
			\hline
			0.235 & 0.242 & 0.269 & 0.291 &  \bf{0.324}\\
			\hline
		\end{tabular}
	\end{center}
	\caption{Ablation study of our full model on SID Sony with each component in terms of H-Mean. A: attention map; E: edge map; M: MS-SSIM; T: text detection loss.}
	\label{table:map_loss_ablation}
    \vspace{-3mm}
\end{table}

\begin{table}[!ht]
	\begin{center}
		\begin{tabular}{c c c c c c}
 			\hline
 			baseline & w/o A & w/o E & w/o M & w/o T & full model\\
 			\hline
 			0.235 & 0.308 & 0.302 & 0.304 & 0.307 & \bf{0.324}\\
 			\hline
 		\end{tabular}
 	\end{center}
 	\caption{Ablation study of our model on SID Sony without each component one at a time in terms of H-Mean. A: attention map; E: edge map; M: MS-SSIM; T: text detection loss.}
 	\label{table:map_loss_ablation_without}
\end{table}


\begin{table}[!t]
	\begin{center}
		\begin{tabular}{l|ccc}
			\hline
			Model & CRAFT & PAN \\
			\hline
			EnlightenGAN~\cite{Jiang2019EnlightenGANDL} & 0.205 & 0.146\\
			Pix2pix~\cite{isola2017image} & 0.281 & 0.224\\
			Ours & \bf{0.348} & \bf{0.278}\\
			\hline
		\end{tabular}
	\end{center}
	\caption{H-Mean on SID Sony when trained on SID Sony and synthetic ICDAR15 datasets.} 
	\label{table:mix_data_ablation}
\end{table}
Also, we test our model further by training it with a mixture of real (SID Sony) and synthetic low-light (ICDAR15) datasets to study if synthetic data can help to improve the image enhancement result. The top 3 models on SID Sony were selected in this experiment. Results in Table \ref{table:mix_data_ablation} showed that there is a significant H-Mean increment when the models are evaluated on the real (SID Sony) test set. We observe that our model gains H-Mean score of 0.024 using CRAFT and 0.012 for PAN. This shows that synthetic low-light data is able to fill up the gap caused by the scarcity of real labeled low-light images and justifies the creation of a synthetic ICDAR15 dataset for such purpose.


\section{Conclusions}
This paper presents an attention-guided and edge-awareness CNN model for extremely low-light image enhancement, particularly on the presence of scene texts. The incorporated self-regularized attention map is proven effective for guiding the network to focus on the dark regions. 
The predicted edge map given the attention map is quantitatively helpful for recovering abundant textures and sharp edges. Most importantly, we propose a novel text detection loss that helps the network attend to those scene text regions to well-recover the scene texts. The experimental results demonstrate that the proposed method consistently outperforms state-of-the-art methods, including low-light image enhancement models and GAN-based ones, in terms of image restoration, text detection, and text spotting on challenging SID Sony and ICDAR15 datasets. Lastly, we also showed that our proposed model trained with additional synthetic extremely low-light images can achieve better results on genuine ones.

\section{Future Directions}
Extending our proposed model to real-time would be beneficial to real-life applications such as collision avoidance of self-driving cars at nighttime and other object detection applications under extremely low-light settings. Besides, we would also like to create a larger dataset that is able to cover more scenes and texts with a wider variety of styles. We believe that these research directions would bring enormous value to both the industrial and research communities.

\clearpage
\balance
\bibliographystyle{IEEEtran}
\bibliography{egbib}

\begin{thebibliography}{10}
\providecommand{\url}[1]{#1}
\csname url@samestyle\endcsname
\providecommand{\newblock}{\relax}
\providecommand{\bibinfo}[2]{#2}
\providecommand{\BIBentrySTDinterwordspacing}{\spaceskip=0pt\relax}
\providecommand{\BIBentryALTinterwordstretchfactor}{4}
\providecommand{\BIBentryALTinterwordspacing}{\spaceskip=\fontdimen2\font plus
\BIBentryALTinterwordstretchfactor\fontdimen3\font minus
  \fontdimen4\font\relax}
\providecommand{\BIBforeignlanguage}[2]{{%
\expandafter\ifx\csname l@#1\endcsname\relax
\typeout{** WARNING: IEEEtran.bst: No hyphenation pattern has been}%
\typeout{** loaded for the language `#1'. Using the pattern for}%
\typeout{** the default language instead.}%
\else
\language=\csname l@#1\endcsname
\fi
#2}}
\providecommand{\BIBdecl}{\relax}
\BIBdecl

\bibitem{Pizer1987AdaptiveHE}
S.~Pizer, E.~P. Amburn, J.~D. Austin, R.~Cromartie, A.~Geselowitz, T.~Greer,
  B.~T.~H. Romeny, and J.~B. Zimmerman, ``Adaptive histogram equalization and
  its variations,'' \emph{Graphical Models \/graphical Models and Image
  Processing \/computer Vision, Graphics, and Image Processing}, vol.~39, pp.
  355--368, 1987.

\bibitem{elik2011ContextualAV}
T.~Çelik and T.~Tjahjadi, ``Contextual and variational contrast enhancement,''
  \emph{IEEE Transactions on Image Processing}, vol.~20, pp. 3431--3441, 2011.

\bibitem{lee2013contrast}
C.~Lee, C.~Lee, and C.-S. Kim, ``Contrast enhancement based on layered
  difference representation of 2d histograms,'' \emph{IEEE Transactions on
  Image Processing}, vol.~22, no.~12, pp. 5372--5384, 2013.

\bibitem{jobson1997properties}
D.~J. Jobson, Z.-u. Rahman, and G.~A. Woodell, ``Properties and performance of
  a center/surround retinex,'' \emph{IEEE Transactions on Image Processing},
  vol.~6, no.~3, pp. 451--462, 1997.

\bibitem{Jobson1997AMR}
D.~Jobson, Z.~Rahman, and G.~A. Woodell, ``A multiscale retinex for bridging
  the gap between color images and the human observation of scenes,''
  \emph{IEEE Transactions on Image Processing}, vol.~6, no.~7, pp. 965--976,
  1997.

\bibitem{Tao2017LowlightIE}
L.~Tao, C.~Zhu, J.~Song, T.~Lu, H.~Jia, and X.~Xie, ``Low-light image
  enhancement using cnn and bright channel prior,'' in \emph{ICIP}, 2017, pp.
  3215--3219.

\bibitem{Tao2017LLCNNAC}
L.~Tao, C.~Zhu, G.~Xiang, Y.~Li, H.~Jia, and X.~Xie, ``{LLCNN}: A convolutional
  neural network for low-light image enhancement,'' in \emph{VCIP}, 2017.

\bibitem{Lore2017LLNetAD}
K.~G. Lore, A.~Akintayo, and S.~Sarkar, ``{LLNet}: A deep autoencoder approach
  to natural low-light image enhancement,'' \emph{Pattern Recognition},
  vol.~61, pp. 650--662, 2017.

\bibitem{Gharbi2017DeepBL}
M.~Gharbi, J.~Chen, J.~Barron, S.~W. Hasinoff, and F.~Durand, ``Deep bilateral
  learning for real-time image enhancement,'' \emph{ACM Transactions on
  Graphics}, vol.~36, no.~4, pp. 1--12, 2017.

\bibitem{Wei2018DeepRD}
C.~Wei, W.~Wang, W.~Yang, and J.~Liu, ``Deep retinex decomposition for
  low-light enhancement,'' \emph{arXiv preprint arXiv:1808.04560}, 2018.

\bibitem{CycleGAN2017}
J.-Y. Zhu, T.~Park, P.~Isola, and A.~A. Efros, ``Unpaired image-to-image
  translation using cycle-consistent adversarial networkss,'' in \emph{ICCV},
  2017.

\bibitem{Jiang2019EnlightenGANDL}
Y.~Jiang, X.~Gong, D.~Liu, Y.~Cheng, C.~Fang, X.~Shen, J.~Yang, P.~Zhou, and
  Z.~Wang, ``Enlightengan: Deep light enhancement without paired supervision,''
  \emph{ArXiv}, vol. abs/1906.06972, 2019.

\bibitem{isola2017image}
P.~Isola, J.-Y. Zhu, T.~Zhou, and A.~A. Efros, ``Image-to-image translation
  with conditional adversarial networks,'' in \emph{CVPR}, 2017.

\bibitem{zhu2020eemefn}
M.~Zhu, P.~Pan, W.~Chen, and Y.~Yang, ``{EEMEFN}: Low-light image enhancement
  via edge-enhanced multi-exposure fusion network.'' in \emph{AAAI}, 2020.

\bibitem{Shi2019LowlightIE}
Y.~Shi, X.~Wu, and M.~Zhu, ``Low-light image enhancement algorithm based on
  retinex and generative adversarial network,'' \emph{ArXiv}, vol.
  abs/1906.06027, 2019.

\bibitem{Baek2019CharacterRA}
Y.~Baek, B.~Lee, D.~Han, S.~Yun, and H.~Lee, ``Character region awareness for
  text detection,'' in \emph{CVPR}, 2019.

\bibitem{wang2019efficient}
W.~Wang, E.~Xie, X.~Song, Y.~Zang, W.~Wang, T.~Lu, G.~Yu, and C.~Shen,
  ``Efficient and accurate arbitrary-shaped text detection with pixel
  aggregation network,'' in \emph{ICCV}, 2019.

\bibitem{trba2019}
J.~Baek, G.~Kim, J.~Lee, S.~Park, D.~Han, S.~Yun, S.~J. Oh, and H.~Lee, ``What
  is wrong with scene text recognition model comparisons? dataset and model
  analysis,'' in \emph{ICCV}, 2019.

\bibitem{Aster2019}
B.~{Shi}, M.~{Yang}, X.~{Wang}, P.~{Lyu}, C.~{Yao}, and X.~{Bai}, ``{ASTER}: An
  attentional scene text recognizer with flexible rectification,'' \emph{IEEE
  Transactions on Pattern Analysis and Machine Intelligence}, vol.~41, no.~9,
  pp. 2035--2048, 2019.

\bibitem{Chen2018LearningTS}
C.~Chen, Q.~Chen, J.~Xu, and V.~Koltun, ``Learning to see in the dark,'' in
  \emph{CVPR}, 2018.

\bibitem{Karatzas2015ICDAR2C}
D.~Karatzas, L.~G.~I. Bigorda, A.~Nicolaou, S.~Ghosh, A.~D. Bagdanov,
  M.~Iwamura, J.~Matas, L.~Neumann, V.~Chandrasekhar, S.~Lu, F.~Shafait,
  S.~Uchida, and E.~Valveny, ``{ICDAR} 2015 competition on robust reading,'' in
  \emph{ICDAR}, 2015.

\bibitem{Fu2016AWV}
X.~Fu, D.~Zeng, Y.~Huang, X.~Zhang, and X.~Ding, ``A weighted variational model
  for simultaneous reflectance and illumination estimation,'' in \emph{CVPR},
  2016.

\bibitem{Guo2017LIMELI}
X.~Guo, Y.~Li, and H.~Ling, ``Lime: Low-light image enhancement via
  illumination map estimation,'' \emph{IEEE Transactions on Image Processing},
  vol.~26, no.~2, pp. 982--993, 2016.

\bibitem{Ying2017ABM}
Z.~Ying, G.~Li, and W.~Gao, ``A bio-inspired multi-exposure fusion framework
  for low-light image enhancement,'' \emph{arXiv preprint arXiv:1711.00591},
  2017.

\bibitem{Goodfellow2014GenerativeAN}
I.~J. Goodfellow, J.~Pouget-Abadie, M.~Mirza, B.~Xu, D.~Warde-Farley, S.~Ozair,
  A.~C. Courville, and Y.~Bengio, ``Generative adversarial nets,'' in
  \emph{NIPS}, 2014.

\bibitem{xue2020arbitrarily}
M.~Xue, P.~Shivakumara, C.~Zhang, Y.~Xiao, T.~Lu, U.~Pal, D.~Lopresti, and
  Z.~Yang, ``Arbitrarily-oriented text detection in low light natural scene
  images,'' \emph{IEEE Transactions on Multimedia}, 2020.

\bibitem{Liu2019RicherCF}
Y.~Liu, M.-M. Cheng, X.~Hu, J.~Bian, L.~Zhang, X.~Bai, and J.~Tang, ``Richer
  convolutional features for edge detection,'' \emph{IEEE Transactions on
  Pattern Analysis and Machine Intelligence}, vol.~41, pp. 1939--1946, 2019.

\bibitem{arbelaez2010contour}
P.~Arbelaez, M.~Maire, C.~Fowlkes, and J.~Malik, ``Contour detection and
  hierarchical image segmentation,'' \emph{IEEE transactions on pattern
  analysis and machine intelligence}, vol.~33, no.~5, pp. 898--916, 2010.

\bibitem{Wang2003MultiscaleSS}
Z.~Wang, E.~Simoncelli, and A.~Bovik, ``Multiscale structural similarity for
  image quality assessment,'' in \emph{ACSSC}, 2003.

\bibitem{Kingma2015AdamAM}
D.~P. Kingma and J.~Ba, ``Adam: A method for stochastic optimization,''
  \emph{arXiv preprint arXiv:1412.6980}, 2015.

\end{thebibliography}
\end{document}